\documentclass[lettersize,journal]{IEEEtran} 
\usepackage{cite}
\usepackage{mathrsfs}
\usepackage[colorlinks]{hyperref}
\usepackage{amsthm}
\usepackage{etoolbox} \makeatletter \patchcmd{\@makecaption} {\scshape} {} {} {} \makeatother
\usepackage[cmex10]{amsmath}
\usepackage{float}
\usepackage{mathtools}
 \usepackage{siunitx}
\usepackage{array}
\usepackage{eqparbox}
\usepackage{bm}

\usepackage[table,dvipsnames]{xcolor}
\usepackage{multicol,booktabs,tabularx}

\usepackage{cases}
\usepackage{algorithm}
\usepackage{paralist}
\usepackage{algpseudocode}

\usepackage[utf8]{inputenc}
\usepackage[english]{babel}
\usepackage{graphicx}
\usepackage{subfigure}
\usepackage{epstopdf}
\usepackage{epsfig}
\usepackage{amssymb}
\usepackage{array}
\usepackage{multirow}

\usepackage{caption}
\usepackage[numbers,sort&compress]{natbib} 
\begin{document}

\title{Near-Field Positioning for XL-MIMO Uniform Circular Arrays: An Attention-Enhanced Deep Learning Approach}
\author{Yuan Gao,~\IEEEmembership{Member,~IEEE},
Xinyu Guo, 
Han Li, 
Jianbo Du,~\IEEEmembership{Member,~IEEE},
Shugong Xu,~\IEEEmembership{Fellow,~IEEE}
\thanks{This work was in part supported by Shanghai Natural Science Foundation under Grant 25ZR1402148, and in part by the 6G Science and Technology Innovation and Future Industry Cultivation Special Project of Shanghai Municipal Science and Technology Commission under Grant 24DP1501001. 
} 

\thanks{Yuan Gao, Xinyu Guo and Han Li are with the School of Communication and Information Engineering, Shanghai University, China, email: gaoyuansie@shu.edu.cn, guoxinyu@shu.edu.cn and 19821215726@shu.edu.cn.}
\thanks{Jianbo Du is with the School of Communication and Information Engineering, Xi'an University of Posts and Telecommunications, Xi'an, China, e-mail: dujianboo@163.com.}
\thanks{Shugong Xu is with Xi’an Jiaotong-Liverpool University, Suzhou, China, email: shugong.xu@xjtlu.edu.cn.}
}

\maketitle

\begin{abstract}
In the evolving landscape of sixth-generation (6G) mobile communication, multiple-input multiple-output (MIMO) systems are incorporating an unprecedented number of antenna elements, advancing towards Extremely large-scale multiple-input-multiple-output (XL-MIMO) systems. This enhancement significantly increases the spatial degrees of freedom, offering substantial benefits for wireless positioning. However, the expansion of the near-field range in XL-MIMO challenges the traditional far-field assumptions used in previous MIMO models. Among various configurations, uniform circular arrays (UCAs) demonstrate superior performance by maintaining constant angular resolution, unlike linear planar arrays. Addressing how to leverage the expanded aperture and harness the near-field effects in XL-MIMO systems remains an area requiring further investigation. In this paper, we introduce an attention-enhanced deep learning approach for precise positioning. We employ a dual-path channel attention mechanism and a spatial attention mechanism to effectively integrate channel-level and spatial-level features. Our comprehensive simulations show that this model surpasses existing benchmarks such as attention-based positioning networks (ABPN), near-field positioning networks (NFLnet), convolutional neural networks (CNN), and multilayer perceptrons (MLP). The proposed model achieves superior positioning accuracy by utilizing covariance metrics of the input signal. Also, simulation results reveal that covariance metric is advantageous for positioning over channel state information (CSI) in terms of positioning accuracy and model efficiency.
\end{abstract}


%
\IEEEpeerreviewmaketitle

\section{Introduction}

The 6G mobile communication systems aim to provide comprehensive coverage, exceptional performance, and intelligent integration \cite{SSnet2025gao,du2024secure,gao2026sidelink,hu2025MADQN}. These goals require unprecedented capabilities, including peak data rates of terabits per second (Tbps), end-to-end latency below one millisecond, and support for over one million devices per square kilometer \cite{jin2024linformer,jiang2025mimo,gao2026csiextra,du2025blockchain,jiang2025mtca}. Achieving these targets presents significant challenges for spectral efficiency, spatial resolution, and environmental sensing \cite{gao2024c2s,hu2024performance,gao2025stochastic,jiang2025towards,gao2019licensed,jiang2025c2s,gao2023fair,gao2023matching,hu2020fairness,gao2026effective}. Traditional MIMO techniques, limited by smaller antenna arrays, are insufficient to meet the demands of advanced applications such as holographic communication and industrial internet of things (IIoT). To overcome these limitations, XL-MIMO systems, which deploy hundreds or thousands of antenna elements, significantly increase the available spatial degrees of freedom \cite{wu2023multiple,cui2022near,dong2024characterizing,liu2023near}. This makes XL-MIMO a crucial technology for enabling the enhanced communication capacity and advanced sensing capabilities, particularly for positioning applications\cite{an2024near,yongliang2022novel,zhao20256g,zhou2024single,hua2025near,xu2025enhanced}, for instance, a higher angle-resolution can be achieved by using a larger number of antennas \cite{gao2024performance,gao2025enabling}, thereby improving positioning accuracy. The improvement of positioning performance is envisioned to enable many location-based services, such as vehicle navigation, UAV tracking, extended reality (XR), etc \cite{zhu2022intelligent,ye2022se,spinelli2017general,ma2025multi,wang2025enhanced}.

Despite these advantages, the positioning enabled by XL-MIMO exhibits distinct features and needs to be carefully addressed. The significant enlargement of antenna apertures in practical systems invalidates the traditional far-field assumption, which treats signals as plane waves \cite{de2020non,han2023toward}. In this case, the angle and delay (with range inferred accordingly) are decoupled, indicating that angle and delay can be estimated separately or sequentially, which significantly lowers the computational complexity of positioning \cite{dai2025tutorial}. However,  in the near-field region, signals exhibit spherical wavefronts, making wavefront curvature and non-radiative (reactive) components non-negligible \cite{bacci2023spherical,yang2024beam}, resulting in the coupling of angle and delay \cite{dai2025tutorial}. Such coupling, on the one hand, provides the unique opportunity to jointly estimate the angle and delay, and on the other hand, makes the positioning challenging \cite{yuan2024scalable,teng2024near}. Addressing these challenges requires reconstructing channel models based on spherical wave theory and developing positioning algorithms specifically adapted to near-field characteristics \cite{zhi2024performance,chen2024near,chen2023cramer}.

Existing research on near-field positioning mainly focuses on uniform linear array (ULA) and uniform planar array (UPA). The ULA \cite{liu2020massive}, characterized by a one-dimensional linear arrangement of antenna elements,offers a simple structure and efficient beamforming, making it suitable for one-dimensional positioning. However, its limited angular resolution and restricted spatial diversity constrain its capability for accurate three-dimensional positioning, thereby limiting its applicability in complex environments. To overcome these limitations, the UPA expands its positioning capability to three dimensions by employing a two-dimensional grid of antenna elements\cite{lu2020omnidirectional}. While this configuration enhances spatial resolution and supports multi-dimensional positioning, the UPA still exhibits angle-dependent performance.Specifically, positioning accuracy for user equipments (UEs) located in front of the array is significantly higher than that for UEs located at its side, due to the variation in effective aperture with respect to the incident angle.

To achieve equal angle resolution and improve positioning performance, the UCA, which features a circular arrangement, has been proposed for wireless positioning \cite{xue2025impact}. UCA provides omnidirectional coverage and robust angle-of-arrival estimation, ideal for scenarios requiring uniform performance across all directions  \cite{wang2023structure,ji2016near,liu2024low,xie2023near}. However, the angle-range coupling of UCA is much more significant than that of ULA or UPA in the near-field region, making the high-accuracy positioning of UCA even more challenging \cite{jackson20222d}. Conventional approaches struggle to cope with this strong coupling. Subspace-based methods, such as multiple signal classification (MUSIC) \cite{salem2024positioning} and estimation of signal parameters via rotational invariance techniques (ESPRIT), suffer from high computational complexity \cite{huang2025sensing,zhi2007near,guzey2015localization,zuo2020subspace}. In contrast, spherical-wavefront-simplification-based methods attempt to reduce the coupling by simplifying the propagation model, which leads to performance degradation due to model mismatch with the actual propagation environment \cite{starer1994passive,huang2002near,liu2014two,he2021mixed}. Artificial intelligence (AI) has therefore been regarded as a promising approach to address angle–range coupling in near-field positioning and has emerged as a key enabler for near-field localization\cite{fabiani2025one,wang2025near,zhang2024deep,zhao2024nflnet,jiang2022deep,liu2024low,cha2023wireless}. Existing AI-based near-field positioning methods typically learn the mapping between the received near-field signals and the source position coordinates without explicitly modeling or simplifying the angle–range coupling. Nevertheless, state-of-the-art approaches still exhibit limited positioning performance, as they primarily rely on local feature extraction or independently process each channel \cite{zhang2024deep,liu2024low}, which prevents them from fully capturing dynamic inter-channel correlations and global spatial dependencies \cite{cha2023wireless}.

To tackle the above challenges, this paper proposes a novel channel-spatial-attention-enhanced deep learning model tailored for near-field UCA positioning. The primary contributions of this research are outlined as follows:

\begin{itemize}

\item As existing AI-driven positioning algorithms (e.g., CNN, MLP, NFLnet \cite{liu2024low}, and ABPN \cite{zhang2024deep}) are generally based on local feature extraction or isolated processing of individual channels, they fail to effectively cope with angle–range coupling. We propose a channel-level attention module, which achieves the "soft" selection of the feature channels. The channel-level attention module is designed to be dual-path to balance the effectiveness of the channel-level feature fusion and computational complexity.

\item
To further capture the long-range spatial dependencies among UCA antenna elements, which are generally insufficiently exploited in existing research, we propose a spatial-level attention module to effectively highlight discriminative regions for positioning, thereby improving positioning performance. A dual-path module integrating average and max pooling is developed to effectively capture the spatial-level features. 
\item Our extensive simulations illustrate that the proposed attention-enhanced deep learning method significantly surpasses existing benchmarks, including the ABPN \cite{zhang2024deep}, NFLnet \cite{liu2024low}, CNN, and MLP. The proposed model achieves superior positioning accuracy by utilizing covariance metric of the input signal. Futhermore, simulation results reveal that the covariance metric is advantageous for positioning over CSI in terms of positioning accuracy and model efficiency.
\end{itemize}

The remainder of this paper is organized as follows. The related work on near-field positioning is comprehensively reviewed in Section II, where the limitations of existing research on near-field UCA positioning are also discussed. In Section III, we formulate the near-field UCA positioning problem and emphasize the unique challenges arising from angle–range coupling. In Section IV, we elaborate on the proposed attention-enhanced deep learning framework and present our main contributions, including the channel-level and spatial-level attention modules. Extensive simulation results with in-depth analysis are presented in Section V. Finally, Section VI concludes this paper.

\section{Related works}
As communication demands continue to escalate, XL-MIMO has emerged as a key technological foundation on the path toward 6G \cite{puspitasari2023emerging}. The substantially enlarged antenna aperture extends the near-field region, making its impact on user signal characteristics increasingly significant and highlighting the need to revisit positioning techniques under this new XL-MIMO paradigm. This section presents a systematic overview of near-field positioning methods. Although prior studies have offered valuable insights, most of them are confined to specific propagation models, array configurations, or individual parameter estimation tasks. A summary of representative works is provided in Table~\ref{tab:survey_summary}. In contrast, this paper aims to deliver a more comprehensive and integrated analysis, covering the fundamental principles of near-field propagation, the influence of different array geometries on positioning performance, and the evolving transition from far-field to near-field positioning

\begin{table*}[t]
\centering
\caption{Summary of Representative Positioning-Related Survey Papers}
\label{tab:survey_summary}
\begin{tabular}{|p{0.7cm}|p{16.6cm}|}
\hline
\textbf{Ref.} & \textbf{Summary} \\
\hline

\cite{italiano2023tutorial} &
(a) Reviews the evolution of cellular-based positioning from early generations to 5G and summarizes 3GPP-defined positioning functionalities.  

(b) Provides simulation results under 3GPP Rel-16 configurations for outdoor and indoor environments.  

(c) Analyzes limitations of current 5G positioning and outlines future research challenges. \\
\hline

\cite{umer2025reconfigurable} &
(a) Summarizes fundamental principles of RIS-assisted positioning, including reflection control and geometry-dependent signal shaping.

(b) Analyzes hardware architectures and deployment strategies such as passive, hybrid, and active RIS.  

(c) Discusses key challenges including near-field modeling, RIS–algorithm joint optimization, and hardware constraints. \\
\hline

\cite{fischer2025systematic} &
(a) Summarizes the principles of angular-based indoor positioning (AoA, AoD, multi-dimensional direction sensing).  

(b) Provides comparative analysis of Wi-Fi, Bluetooth, UWB, mmWave arrays, acoustic, and optical sensing platforms.  

(c) Discusses limitations related to multipath, near-field effects, array calibration, and feature fusion. \\
\hline

\cite{chen2022tutorial} &
(a) Introduces the physical principles of THz-band positioning (ultra-wide bandwidth, directionality, multipath). 

(b) Describes ToA/ToF, AoA, and joint delay–angle estimation enabled by THz arrays.  

(c) Summarizes challenges such as molecular absorption, near-field modeling for ultra-large arrays, and synchronization issues. \\
\hline

\cite{shastri2022review} &
(a) Describes principles of device-based and device-free mmWave sensing/positioning.  

(b) Reviews radar-style processing, compressive sensing, and learning-based recognition methods.  

(c) Summarizes challenges including blockage robustness, hardware constraints, and near-field sensing. \\
\hline

\cite{zhu2020indoor} &
(a) Summarizes principles of indoor fingerprinting (feature construction, radio-map generation).  

(b) Reviews RSS, CSI, magnetic, visual, and deep learning–based fingerprinting techniques.  

(c) Discusses key challenges such as environmental dynamics, scalability, calibration overhead, and near-field effects. \\
\hline

our paper &
(a) The motivation for developing XL-MIMO in 6G systems is introduced, including the ultra-high spatial resolution enabled by extremely large arrays, the emergence of strong near-field spherical-wave effects, and the breakdown of traditional far-field assumptions.

(b) The impact of different antenna array configurations on positioning is discussed, covering linear, planar, and distributed arrays with respect to resolution capability, coverage characteristics, and depth perception.

(c) The evolution from conventional far-field positioning to near-field positioning algorithms is summarized, highlighting challenges such as accurate near-field channel modeling, high-dimensional parameter estimation, and the growing computational burdens introduced by ultra-large arrays. \\
\hline

\end{tabular}
\end{table*}

\subsection{XL-MIMO Assisted Positioning}

XL-MIMO has been a key enabler for 6G to provide substantially higher spatial degrees of freedom, narrower beams, and stronger user discrimination capability \cite{liu2025sensing}. As the array aperture extends to meter-level or even larger dimensions, the underlying propagation characteristics undergo fundamental changes. Traditional MIMO systems typically assume far-field plane-wave propagation, where the distance differences between the user and individual antenna elements are negligible, allowing the impinging wavefront to be modeled as planar. However, this assumption no longer holds in XL-MIMO: noticeable distance variations may arise across different regions of the array, resulting in a curved electromagnetic wavefront and thereby placing the user within the near-field region of the array \cite{tang2025revisiting}. Under near-field propagation, the received wavefront at the array is better modeled as spherical, and both the phase and amplitude across antenna elements depend on the user’s three-dimensional position \cite{yin2017scatterer,zhou2015spherical}.

Near-field effects exert a significant impact on positioning performance. In principle, far-field positioning primarily relies on the angle of arrival (AoA), making it difficult for a single array to extract precise range information \cite{sesyuk2022survey}. In contrast, near-field propagation introduces substantial distance variations across antenna elements, embedding depth information directly into the phase structure of the received signal and effectively adding an additional geometric constraint. As a result, the positioning process transitions from angle-based two-dimensional estimation to joint angle–range three-dimensional reconstruction, substantially enhancing 3D positioning resolution and improving the identifiability of parameters such as AoA and ToA. Furthermore, the enhanced spatial focusing capability of XL-MIMO in the near field provides an additional physical gain, offering new opportunities for achieving high-precision positioning in future 6G systems.

Typical XL-MIMO architectures can be classified into linear arrays, planar arrays and circular arrays. Linear arrays offer simple deployment and calibration, making them highly practical for initial prototypes and scenarios with linear signal propagation \cite{wu2023enabling,wang2016mixed,tang2023near}. They are easily scalable, as adding more antennas directly enhances AoA accuracy and enriches channel state information. However, linear arrays suffer from poor angular resolution in endfire directions, i.e., the directions along the axis of the array, reducing effectiveness in certain orientations. They also struggle with resolving multipath in three-dimensional spaces, leading to lower overall precision. Planar arrays extend the aperture to two dimensions, improving angular resolution in both azimuth and elevation and supporting more accurate 3D positioning and beam control \cite{cui2023near}. However, planar arrays also suffer from poor angular resolution in endfire directions, limiting their positioning capabilities in teh corresponding directions \cite{xie2023near,chen2025near}. Circular arrays, by uniformly distributing elements along a circular perimeter, achieve full rotational symmetry and offer directionally consistent sampling of near-field wavefronts. This symmetry strengthens the identifiability of parameters such as AoA and range, and enables more balanced spatial resolution across directions \cite{liu2024low,sun2023efficient,wang2023structure}. Assumption of far-field plane-wave propagation is not valid in near-field positioning; therefore, specific positioning algorithms for near-field have to be developed. Existing positioning techniques broadly fall into two categories: model-based methods \cite{xia2025near,li2024near,rahal2024ris,wu2025hybrid,wu2020rank} and AI-driven approaches \cite{park2025towards,macias2024ris,jiang2025near}.

\subsection{Model-driven Approaches for Near-Field Positioning}
Model-driven near-field positioning approaches can be classified into subspace-based schemes and spherical-wavefront simplification–based schemes.
\subsubsection{Subspace-based schemes}
Subspace-based schemes rely on subspace decomposition techniques, such as the MUSIC algorithm \cite{salem2024positioning} and ESPRIT. Standard 2D-MUSIC for near-field sources requires an exhaustive global 2D spectral search over both range and bearing parameters, which is computationally intensive and susceptible to spurious local peaks. Huang et al. \cite{huang2002near} extended traditional MUSIC, originally designed for far-field sources (1D DOA estimation under the plane-wave assumption), to the near-field spherical wavefront model by enabling joint estimation of range and bearing for multiple sources. It constructs a 2D MUSIC spectrum by searching over a 2D grid of (range, bearing) parameters, using the noise subspace to identify peaks corresponding to source locations. This handles wavefront curvature explicitly, allowing passive localization of multiple near-field narrow-band sources with a uniform linear array, though it incurs high computational cost due to the full 2D search. This limitation is mitigated in \cite{starer1994passive}, which proposes a modification that minimizes the MUSIC cost function subject to geometrical constraints imposed by wavefront curvature. It employs a modified path-following method that reduces the global 2D search to $2(M-1)$ independent 1D searches (where $M$ is the number of elements in a uniform linear array). This enables high parallelism, global convergence, and significantly improved computational efficiency while maintaining high accuracy for multiple narrow-band near-field sources. Huang et al. \cite{huang2025sensing} propose a low-complexity variant of 2D-MUSIC to estimate the angle and range for mobile targets in the near field, reducing exhaustive searches while maintaining accuracy in dynamic environments. \cite{zhi2007near} applies generalized far-field ESPRIT for DOA, followed by 1D search for range, reducing complexity from 2D to 1D MUSIC-like searches.

To improve the robustness to colored and nonuniform noise, \cite{guzey2015localization} extends MUSIC with techniques (e.g., prewhitening or covariance adjustments) to handle spatially colored noise, improving robustness for unintended emitters, \cite{zuo2020subspace} uses Toeplitz-like matrix reconstruction from anti-diagonal covariance elements to convert nonuniform noise to uniform, enabling standard MUSIC application with high resolution. 

For mixed near-field and far-field positioning, a two-stage MUSIC variant based on higher-order (typically fourth-order) cumulants has been proposed to handle scenarios with simultaneous near-field and far-field sources, avoiding full 2D searches and parameter pairing issues. Liang et al. \cite{liang2009passive} construct two specialized fourth-order cumulant matrices: one for DOA estimation of all sources via MUSIC, and another for joint DOA–range estimation. They further perform two separate 1D MUSIC spectral searches to localize far-field (DOA only) and near-field (DOA and range) sources, alleviating aperture loss and avoiding 2D searches. Liu et al. Liu et al. \cite{liu2014two} exploit the Toeplitz structure of the far-field covariance matrix through a two-stage differencing operation to eliminate far-field components. MUSIC is then applied to the residual matrix for near-field DOA and range estimation, enabling automatic source classification without prior knowledge of source types.

\subsubsection{Spherical-wavefronts-simplification-based schemes}
The second type of model-driven approaches are based on the simplification of the spherical wavefront through  array partitioning or approximations. Array partitioning-based schemes divide the extremely large array into subarrays or regions to enable localized far-field or simplified models. For instance, Starer and Nehorai \cite{starer1994passive} incorporated distance information into steering matrices, adapting the MUSIC algorithm for near-field scenarios. Later efforts by Huang and Barkat \cite{huang2002near} improve computational efficiency by decoupling the two-dimensional MUSIC algorithm into separate one-dimensional angle and range estimation stages.
 Alternatively, the MILE framework \cite{he2021mixed} employed high-precision channel models to directly perform geometric positioning of mixed sources, eliminating the need for explicit signal separation or source classification. Additionally, recent work by Teng et al. \cite{teng2024near} introduced array partitioning in XL-MIMO systems, using message-passing algorithms applied to array subgroups for positioning. Liu et al. \cite{liu2024low} proposed an efficient near-field localization algorithm tailored for XL-MIMO systems, utilizing sectored uniform circular arrays. This design exploits azimuthal consistency in resolution and enables FFT-accelerated processing for low computational complexity, delivering high localization accuracy with significantly reduced runtime compared to conventional algorithms in massive antenna scenarios.

Approximation-based schemes mathematically simplify the spherical wavefronts using Taylor expansions, Fresnel, piecewise far-field, or chirp models. Xi et al. \cite{xi2025near} proposed a second-order Taylor approximation of spherical wavefronts to decompose steering vectors into far-field and chirp-modulated components, enabling a low-dimensional discrete chirp rate subspace for gridless angle-range recovery through atomic norm minimization for channel estimation, along with theoretical error bounds that simplify positioning in sparse multipath without full polar grids. Chen et al. \cite{chen2025near} developed a fractional Fourier transform-based near-field beamspace that partitions into high/low mainlobes and sidelobes to simplify spherical wavefronts, where high mainlobe approximations (such as 2D Gaussian) enable precise angle-distance estimation and improve positioning resolution over traditional methods.

\subsection{AI-driven Approaches for Near-Field Positioning}

AI-based near-field positioning predominantly relies on deep learning models, such as DNNs and CNNs \cite{OTFSGao2025}. Gast et al. \cite{gast2025near} introduced AI-augmented subspace techniques, including a deep neural network-enhanced 2D MUSIC algorithm that learns surrogate covariance matrices and a cascaded decoupling method for angle and range estimation. These methods enhance robustness to real-world impairments such as coherent sources, array miscalibrations, and limited snapshots, while preserving interpretability and reducing complexity. \cite{jiang2022deep} proposed an end-to-end deep residual learning network that directly regresses near-field source locations from multi-dimensional array covariance inputs. This approach effectively handles unknown spatially colored noise without prior assumptions, achieving robust and accurate localization in challenging noise environments.  \cite{zhao2024nflnet} proposed NFLnet, a dedicated deep neural network architecture optimized for near-field vehicle localization in practical scenarios. It addresses nonlinear positioning challenges in device-free settings, demonstrating superior accuracy and efficiency for road vehicle surveillance compared to traditional methods. \cite{jang2024dnn} jointly optimized beamforming and positioning through multi-task neural network training. Notably, many current neural network implementations depend heavily on simulated data rather than real-world measurements, with limited publicly available datasets. Among these, MaMIMO \cite{li2021toward} stands out as a prominent dataset, offering dense indoor CSI measurements for fine-grained positioning tasks. Advanced deep-learning models, such as attention mechanisms \cite{jin2024efficient,jin2025dual}, have been proposed for positioning. \cite{zhang2024deep} proposed an ABPN to utilize the rich features of CSI for positioning enhancement via channel and spatial attention modules. In order to solve the problem of near-field AoA estimation, Wang et al. \cite{wang2025near} propose complex convolutional Kolmogorov-Arnold network (CCKAN). This model introduces complex convolutional networks into the Kolmogorov-Arnold network (KAN), which can make full use of complex-valued signals as input and avoid information loss caused by dividing signals into irrelevant real and imaginary parts.In the context of the hybrid analog-digital beamformer, \cite{fabiani2025one} utilized a CNN to perform the design of the analog beamformer and simultaneously estimate the near-field position of a single user in a single snapshot. \cite{khan2025drl} proposed a multi-point positioning scheme (MLS) based on deep reinforcement learning (DRL) for efficiently locating Internet of Things (IoT) devices using a single autonomous flying vehicle (AAV) equipped with a large-scale multi-antenna configuration, achieving positioning in Near Field Communication (NFC). Zhu et al. \cite{zhu2025bio} proposed a bio-inspired dendritic liquid neural network (DLNN) achieving improved accuracy, robustness, and low-latency inference through multi-branch dendritic processing and liquid-state temporal modeling.

\section{System model and problem formulation}
\label{sec:system_modeling}

As shown in Figure \ref{fig:uca-geometry}, we consider a wireless positioning system consisting of a base station (BS) equipped with a massive MIMO UCA  and a single antenna user equipment (UE). The UCA antenna array consists of $N$ antenna elements distributed uniformly along a circular arc with radius $R$ m. The array center is located at the origin of the polar system and the coordinates of the $n$-th array element are given by:
\begin{equation}
\begin{bmatrix}
\theta_n \\
r_n 
\end{bmatrix} 
= 
\begin{bmatrix}
\frac{\pi}{6} + \frac{(n-1)\pi}{3(N-1)} \\
R 
\end{bmatrix},
\end{equation}

The single-antenna UE is located at the near-field region of the BS with spherical coordinates $(\eta_s, r_s)$. We consider an uplink positioning scenario, where received signal at the BS at the $k$-th time step is expressed as:

\begin{figure}[htbp]
  \centering
  \includegraphics[height=0.25\textheight]{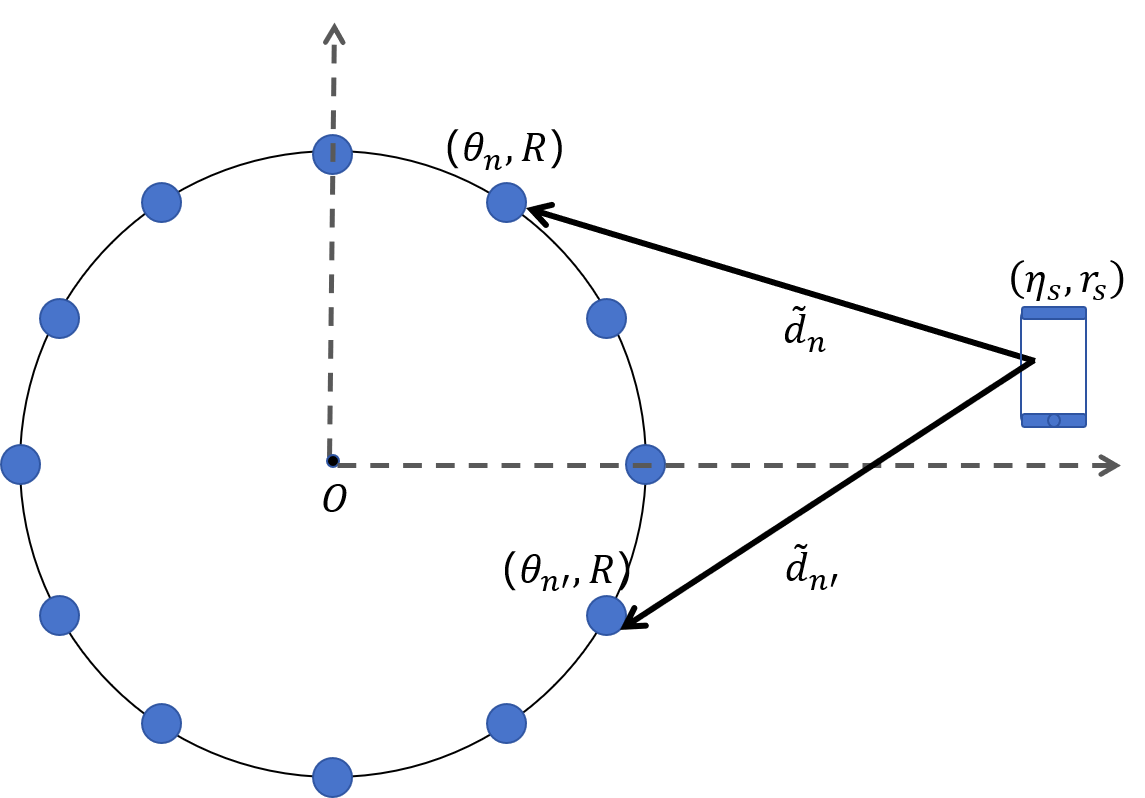} 
  \caption{Geometric relationship between UCA and UE in near-field region.}
  \label{fig:uca-geometry}
\end{figure}

\begin{equation} 
\mathbf{y}_k = \mathbf{C}\mathbf{h}_k \mathbf{s}_k + \mathbf{n}_k,
\end{equation}where $\mathbf{s}_k$ and $\mathbf{n}_k\in \mathcal{CN}(0,\sigma^2)$ denotes the positioning signal and the additive Gaussian noise (AWGN) at the $k$-th time step. $\mathbf{h}_k\in \mathbb{C}^{N\times 1}$ is the uplink channel vector at the $k$-th time step. Since UE is located within the near-field region of the BS, the conventional far-field plane-wave assumption is no longer valid. Instead, $\mathbf{h}_k\in \mathbb{C}^{N\times 1}$ is modeled using spherical-wave:

\begin{equation}\label{channel_matrix}
\mathbf{h}_k=\begin{pmatrix}
\frac{1}{d_1}e^{-j\frac{2\pi}{\lambda}\widetilde{d}_1}\\
 ...\\
\frac{1}{d_N}e^{-j\frac{2\pi}{\lambda}\widetilde{d}_N}
\end{pmatrix},
\end{equation}where $\lambda$ is the wavelength of the carrier frequency. $\widetilde{d}_n$ is the distance between the UE and $n$-th antenna of the BS, which is expressed as:
\begin{equation}\label{distance_angle}
    \widetilde{d}_n=\sqrt{r^2_s+r^2_n-2r_sr_n\text{cos}(\eta_s-\theta_n)}.
\end{equation}

The positioning problem in the above system is using the received signal of multiple time steps to acquire the position of the UE, i.e.,:
\begin{equation}\label{Pos_problem}
    (\widetilde{\eta}_s, \widetilde{r}_s)=\mathcal{P}([\mathbf{y}_1,...,\mathbf{y}_K]),
\end{equation}where $\widetilde{\eta}_s$ and $\widetilde{r}_s$ are the estimated angle and range of the UE in the polar coordinates. $\mathcal{P}(*)$ is the positioning function that estimates the positioning of input signal $*$. As demonstrated in Eq. (\ref{channel_matrix}) and (\ref{distance_angle}), $(\widetilde{\eta}_s, \widetilde{r}_s)$ are coupled nonlinearly, making the design of the positioning function $\mathcal{P}(*)$ for UCA much more challenging than that for ULA or UPA. In contrast, the angle–range coupling in ULA or UPA can be approximately simplified to a linear form.

\begin{figure}[htbp]
  \centering
  \includegraphics[height=0.22\textheight]{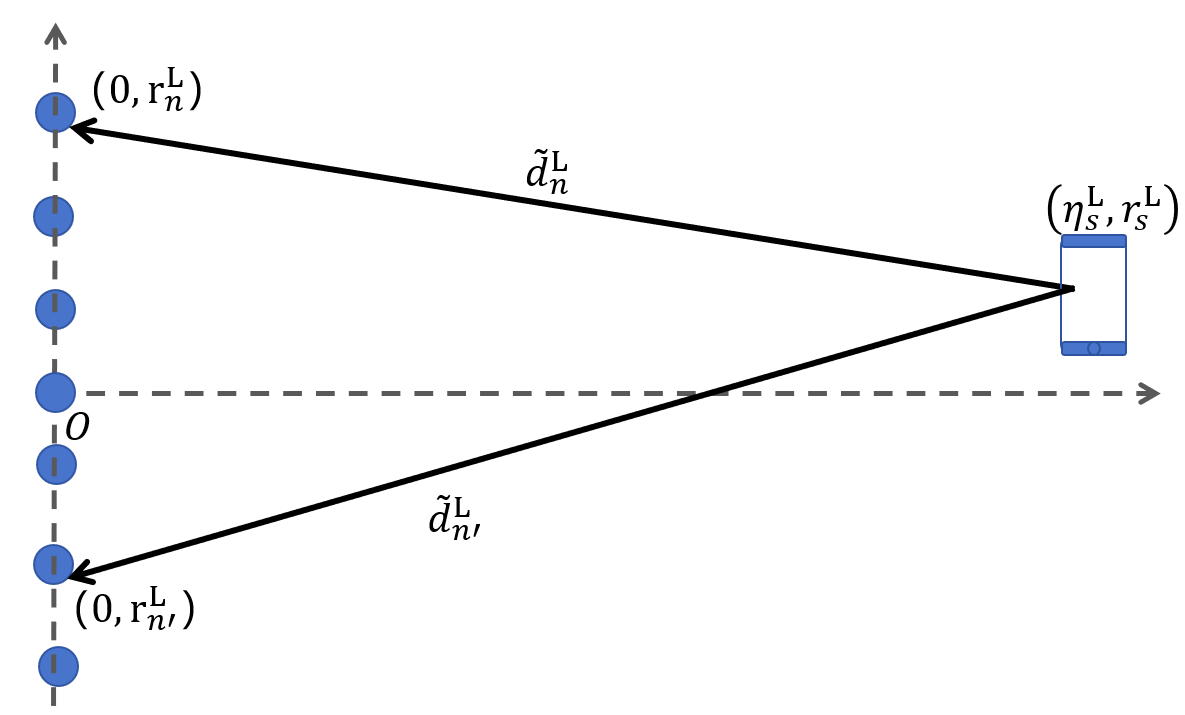} 
  \caption{Geometric relationship between ULA and UE in near-field region.}
  \label{fig:ula-geometry}
\end{figure}
\begin{figure*}[t]  
\centering
\includegraphics[width=\textwidth]{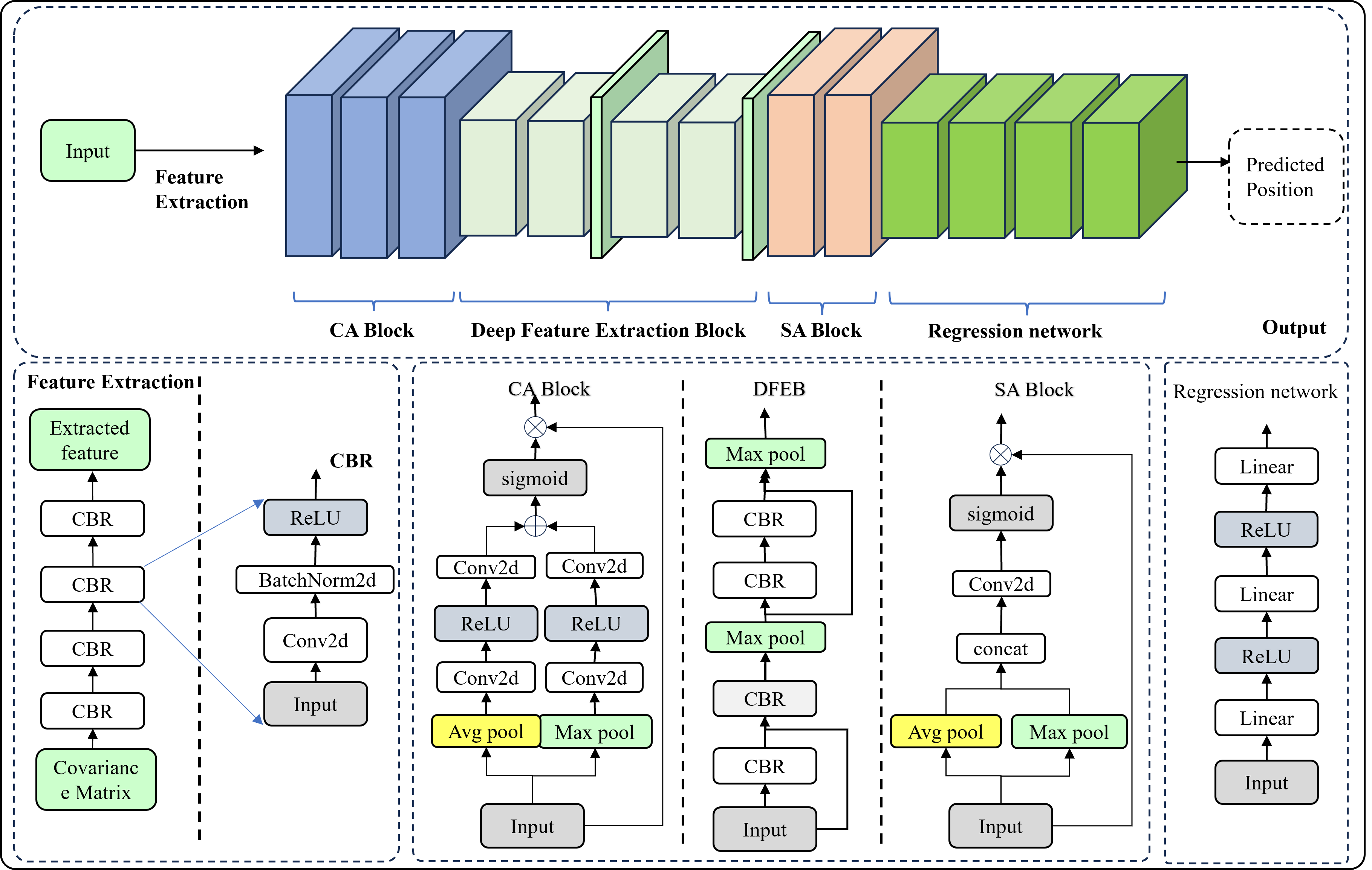} 
\caption{Architecture of the proposed model}
\label{fig:model}  
\end{figure*}

In the following section, we will illustrate the linear angle-range coupling in ULA as an example, similar analysis can be extended to UPA. As shown in Fig. \ref{fig:ula-geometry}, the distance $\widetilde{d}^{\text{L}}_n$ between the UE and $n$-th antenna of the BS is expressed as:
\begin{equation}\label{distance_angle_ULA}
\widetilde{d}^{\text{L}}_n=\sqrt{{(r^{\text{L}}_s)^2}+(r^{\text{L}}_n)^2-2r^{\text{L}}_sr_n^{\text{L}}\text{cos}(\eta_s^{\text{L}})},
\end{equation}the phase difference between the $n$-th antenna element and the antenna element at the origin $O$ is proportional to the propagation path difference $\widetilde{d}^{\text{L}}_n-r^{\text{L}}_s$, which is expressed mathematically as: 
\begin{equation}\label{Phase_diff}
\widetilde{d}^{\text{L}}_n-r^{\text{L}}_s=\sqrt{{(r^{\text{L}}_s)^2}+(n\Delta)^2-2r^{\text{L}}_s(n\Delta)\text{cos}(\eta_s^{\text{L}})}-r^{\text{L}}_s
\end{equation}

The near field region is defined by the Fresnel region equation as: 
\begin{equation}\label{Fresnel_ineq}
 0.62\sqrt{\frac{D^3}{\lambda}}\le r_s^\text{L} \le \frac{D^2}{\lambda},  
\end{equation}considering the left part of the inequality in Eq. (\ref{Fresnel_ineq}), we have:
\begin{align}
&0.3844D^3\le (r_s^\text{L})^2\lambda,\\
&\xrightarrow{a} 0.3844(N\Delta/2)^3\le (r_s^\text{L})^2\lambda,\\
&\xrightarrow{b} 0.024025N(N\Delta)^2\le (r_s^\text{L})^2,\\
&\xrightarrow{c} \left( \frac{N\Delta}{r_s^\text{L}} \right)^2\le\frac{41.6}{N}<<1,
\\&\xrightarrow{d} \frac{n\Delta}{r_s^\text{L}} \le\frac{N\Delta}{r_s^\text{L}}<<1\label{ineq_final},
\end{align}where transformation ($a$) is achieved by assuming $D\cong N\Delta/2$; transformation ($b$) is achieved by assuming $\Delta\cong\lambda/2$; transformation ($c$) is achieved due to the fact that the number of antenna elements of an XL-MIMO is generally of the level of thousands; transformation ($d$) is achieved because $n<=N$ 
   
By using Eq. (\ref{ineq_final}) ($n\Delta/{r_s^\text{L}}<<1$) in Eq. (\ref{Phase_diff}) to perform Taylor expansion, we have:
\begin{align}
\widetilde{d}^{\text{L}}_n-r^{\text{L}}_s&\cong -n\Delta\text{cos}(\eta_s^{\text{L}})+\frac{(n\Delta)^2}{2r^{\text{L}}_s},\\
&=-n\Delta\text{cos}(\eta_s^{\text{L}})+o(n\Delta)\label{Ilear_ill},    
\end{align}where Eq. (\ref{Ilear_ill}) indicates that the range $r^{\text{L}}_s$ and the angle $\eta_s^{\text{L}}$ are approximately linearly-coupled, which makes the range and angle estimation of ULA or UPA easier than that of the UCA.

\section{Proposed model}
\subsection{Model architecture}As illustrated in Fig. \ref{fig:model}, we propose a deep learning-based framework for UCA near-field positioning that effectively copes with the angle–range coupling. We first design an initial feature extraction block to increase the feature dimension of the input data, which is beneficial for subsequent processing. To balance the effectiveness of the channel-level feature fusion and computational complexity, we propose a dual-path channel-level attention mechanism, which achieves a "soft" selection of the feature channels via training. Subsequently, a deep feature extraction block is proposed to capture multi-scale feature representations while maintaining feature continuity through residual connections. To establish long-range spatial dependencies and effectively highlight discriminative regions for positioning, a spatial-level attention block is proposed. Finally, the position is estimated via a MLP-based regression network.
\subsubsection{Initial feature extraction block}
The input block serves as the initial feature extractor and contains four convolutional, batch normalization and ReLU (CBR) modules. In the field of deep learning, the CBR module is a common building block composed of three sequential key operations: convolutional layer, batch normalization, and ReLU activation function. The convolutional layer serves to perform core feature extraction, learn spatial features through convolutional kernels, and modify channel dimensions. The convolution operation freely controls the number of output channels by setting out channels, where each output channel corresponds to an independent convolutional kernel that learns distinct features, thereby achieving channel dimension expansion. Batch normalization aims to standardize the output of the previous layer, accelerate training, improve model stability, and reduce sensitivity to initialization. The ReLU activation introduces nonlinearity and enhances the model's expressive capability. 

Through the CBR module in the input block, the data is expanded from 2 channels to 128 channels, and this dimensionality increase enables the network to perform feature learning in a higher-dimensional space. The stacked convolution operations establish rich local connectivity patterns:
\begin{equation}
    F =\sigma \left ( BN\left ( Conv\left ( F_{in}  \right )  \right )  \right ) ,
\end{equation}
where $\sigma $ denotes the ReLU activation, $F_{in} \in\mathbb{R}^{\mathrm {2\times N\times N}   } $ denotes the covariance matrix used as the network input, and $F \in\mathbb{R}^{\mathrm {128\times N\times N}   } $ denotes the output of the input block. The purpose of the input block is to increase the dimensionality of the input data, which can provide optimization space for subsequent channel attention mechanisms.

\subsubsection{Channel attention (CA) block}

Subsequently, we introduce a dual-path channel attention mechanism to effectively fuse the channel-level features. As shown in Fig. 2, the module computes channel-level attention weights through parallel spatial pooling operations:
\begin{equation}
    F_{\mathrm{CA}} = A_{\mathrm {C}} \otimes F,
\end{equation}
where $\otimes $ denotes channel-wise multiplication, $A_{\mathrm {C}}$ denotes the weight of each channel dimension, which can be written as:
\begin{equation}
\begin{aligned}
             &A_{\mathrm {C}} \\&=\mathrm {Sigmoid}\left ( \mathrm {Conv}\left ( \mathrm {Avgpool}\left ( F \right )\right )+  \mathrm {Conv}\left ( \mathrm {Maxpool}\left ( F \right )\right )\right ).
\end{aligned}
\end{equation}

The convolutional layer employs 1×1 convolutions, effectively capturing inter-channel relationships. Channel attention focuses on the overall importance of each channel and aims to suppress noise channels are less relevant or noisy with respect to the target task. We adopt two parallel paths, namely global average pooling (GAP) and global max pooling (GMP) \cite{zhao2024improved,sabri2020comparison}. GAP compresses the spatial dimensions to compute per-channel mean, which reflects global activation strength per channel. GMP captures the most significant feature responses in each channel and focuses on extreme activation values. This dual-path design provides complementary assessment criteria for the importance of channels. The outputs of GAP and GMP are concatenated and processed through 1×1 convolution, and then the channel attention weights between 0 and 1 are generated through the sigmoid function. The learned attention weights are applied to channel multiplication with the original feature F. The important channels will be enhanced and the unimportant channels will be suppressed. The core idea of the attention mechanism is to selectively focus on important information (channels) and ignore secondary information. We recalibrate the feature response through the automatically learned weights, achieving the "soft" selection of the feature channels.

\subsubsection{Deep feature extraction block (DFEB)}

DFEB incorporates residual learning with progressive downsampling. The block alternates between basic residual units and max-pooling layers, implementing the following operations:
\begin{equation}
\begin{aligned}
    &F_{\mathrm {DFE}}\\ &= \sigma \left ( \mathrm {BN}\left ( \mathrm {Conv}\left ( \sigma \left ( \mathrm {BN}\left ( \mathrm {Conv}\left ( F_{\mathrm {CA}}\right )\right )\right )\right )\right )+  \mathrm {S}\left ( F_{\mathrm {CA}}\right )\right ),
\end{aligned}
\end{equation}
where $\mathrm {S\left ( \cdot  \right ) }$ denotes the shortcut connection using identity mapping or 1×1 convolution for dimension matching. Max-pooling with $2\times 2$ window is applied after every two residual units, systematically reducing spatial dimensions from 64×64 to $16\times 16$ through three stages. This hierarchical structure enables multi-scale feature representation while maintaining feature continuity through residual connections.

\begin{table*}[]
\centering
\caption{\\Architecture of the proposed positioning method}
{
\begin{tabular}{c|c|c|c}
\hline
Block & Operation & Parameters & Output Shape \\ \hline
Input & —— & —— & (B,2,64,64) \\ \hline
Input Block & Conv units & $\begin{bmatrix}2,128,3\times3  \\128,128,3\times 3 \\128,128,3\times 3\\128,128,3\times 3
\end{bmatrix}$ & (B,128,64,64) \\ \hline
\multirow{4}{*}{CA Block} & Avg Pooling & Global→1×1 & (B,128,1,1) \\ \cline{2-4} 
 & Max Pooling & Global→1×1 & (B,128,1,1) \\ \cline{2-4} 
 & Conv units & $\begin{bmatrix}128,8,1\times1  \\8,128,1\times1
\end{bmatrix}$ & (B,128,1,1) \\ \cline{2-4} 
 & \begin{tabular}[c]{@{}c@{}}Sigmoid + \\ Element-wise Multiply\end{tabular} & —— & (B,128,64,64) \\ \hline
\multirow{2}{*}{DFEB} & Max pooling & $\left ( 2\times 2,2 \right ) $ & \multirow{2}{*}{(B,128,16,16)} \\ \cline{2-3}
 & BasicBlocks & $\begin{bmatrix}128,128,3\times3\\128,128,3\times3\\128,128,3\times3\\128,128,3\times3\end{bmatrix}$ &  \\ \hline
\multirow{4}{*}{SA Block} & Avg Pooling & dim=1 (channel) & (B,1,16,16) \\ \cline{2-4} 
 & Max Pooling & dim=1 (channel) & (B,1,16,16) \\ \cline{2-4} 
 & Conv units & $\left[2,1,7\times7\right]$ & (B,1,16,16) \\ \cline{2-4} 
 & \begin{tabular}[c]{@{}c@{}}Sigmoid + \\ Element-wise Multiply\end{tabular} & —— & (B,128,16,16) \\ \hline
\multirow{2}{*}{MLP} & Flatten & —— & (B, 32768) \\ \cline{2-4} 
 & Linear & $\begin{bmatrix}32768,128\\128,128\\128,2\end{bmatrix}$ & (B,2) \\ \hline
\end{tabular}
}
\end{table*}

\subsubsection{Spatial attention (SA) block}

SA block processes the enhanced features through spatial-wise attention:
\begin{equation}
    F_{\mathrm {SA}} = A_{\mathrm {S}} \otimes F_{\mathrm {DFE}}, 
\end{equation}where $A_{\mathrm {S}}$ denotes the spatial attention weight,which can be written as:
\begin{equation}
\begin{aligned}
             A_{\mathrm {S}}& \\=&\mathrm {Sigmoid}\left ( \mathrm {Conv}\left (  \left  [ \mathrm {Avgpool}\left ( F_{\mathrm {DFE}}\right);\mathrm {Maxpool}\left ( F_{\mathrm {DFE}}\right ) \right ] \right ) \right ), 
\end{aligned}
\end{equation}where $\left [ \cdot ;\cdot  \right ]   $ denotes channel concatenation. It is worth mentioning that the pooling operation here is carried out along the channel dimension. The convolutional layer uses a large convolutional kernel with a size of $7\times 7$, which can establish long-range spatial dependencies, effectively highlighting discriminative regions for positioning. Compared with channel attention, the SA block generates spatial weight maps by concatenating the average and maximum pooling of channel dimensions. Spatial attention focuses on the sensitivity of pixel-level positions. Spatial attention applies average and max pooling along the channel dimension to achieve cross channel information aggregation. The average pooling along the channel dimension reflects multi-channel consensus, while maximum pooling emphasizes peak activation and has different spatial focusing characteristics. Direct concatenation can preserve the inherent differences between the two statistical patterns and avoid feature blurring caused by addition.\par

\subsubsection{Positioning regression Block}

The final regression module converts high-dimensional features into two-dimensional position predictions, namely the distance and angle, through an MLP\cite{bartol2022linear}:
\begin{equation}
    \hat{y} = \mathrm {W_{n}}\left ( \sigma \left ( \dots \sigma \left ( \mathrm {W_{1} }\left ( \mathrm {Flatten}\left ( F_{\mathrm {SA}}\right )\right )\right )\right )\right ),   
\end{equation}where $\mathrm {W_{i} }$ denotes fully-connected layers. MLP first flattens the feature map of each channel and converts the spatial-channel mixed features into a vector form suitable for processing in fully connected networks through the normalization processing of input layer $\mathrm {W_{1} }$. Then, deep features are extracted through N-2 hidden layers, and finally, two-dimensional position parameters $\hat{y} = \left [\hat{r}, \hat{\eta }\right] ^{T}$ are regressed through output layer $\mathrm {W_{n} }$.

\subsection{Training}

\begin{table}[htbp]
\centering
\caption{System Simulation Parameters}
\label{tab:simulation_params}
\begin{tabular}{lc}
\toprule
\textbf{Parameter} & \textbf{Value} \\
\midrule
Array Type & Uniform Circular Array (UCA) \\
Array Radius (m) & 1 \\
Number of antenna elements & 64 \\
UE Distance Range (m) & (2, 10) \\
UE Azimuth Range (°) & (-30, 150) \\
Carrier Frequency (GHz) & 3.5 \\
SNR of the test scenarios & 0, 20 dB \\
Number of snapshots for input signal & 50, 100 \\
Optimizer & Adam \\
Batch Size & 32 \\
Training samples& 8,000 \\
Testing samples & 2,000 \\
Training Epochs & 200 \\
Learning Rate & 0.0003 \\
Training epochs& 200 \\
Hardware platform & GeForce RTX 4070 GPU \\
\bottomrule
\end{tabular}
\end{table}

\begin{figure*}
    \centering
    \subfigure[Covariance matrix (50 snapshots) as input at 0 dB.]{
        \includegraphics[width=0.48\linewidth]{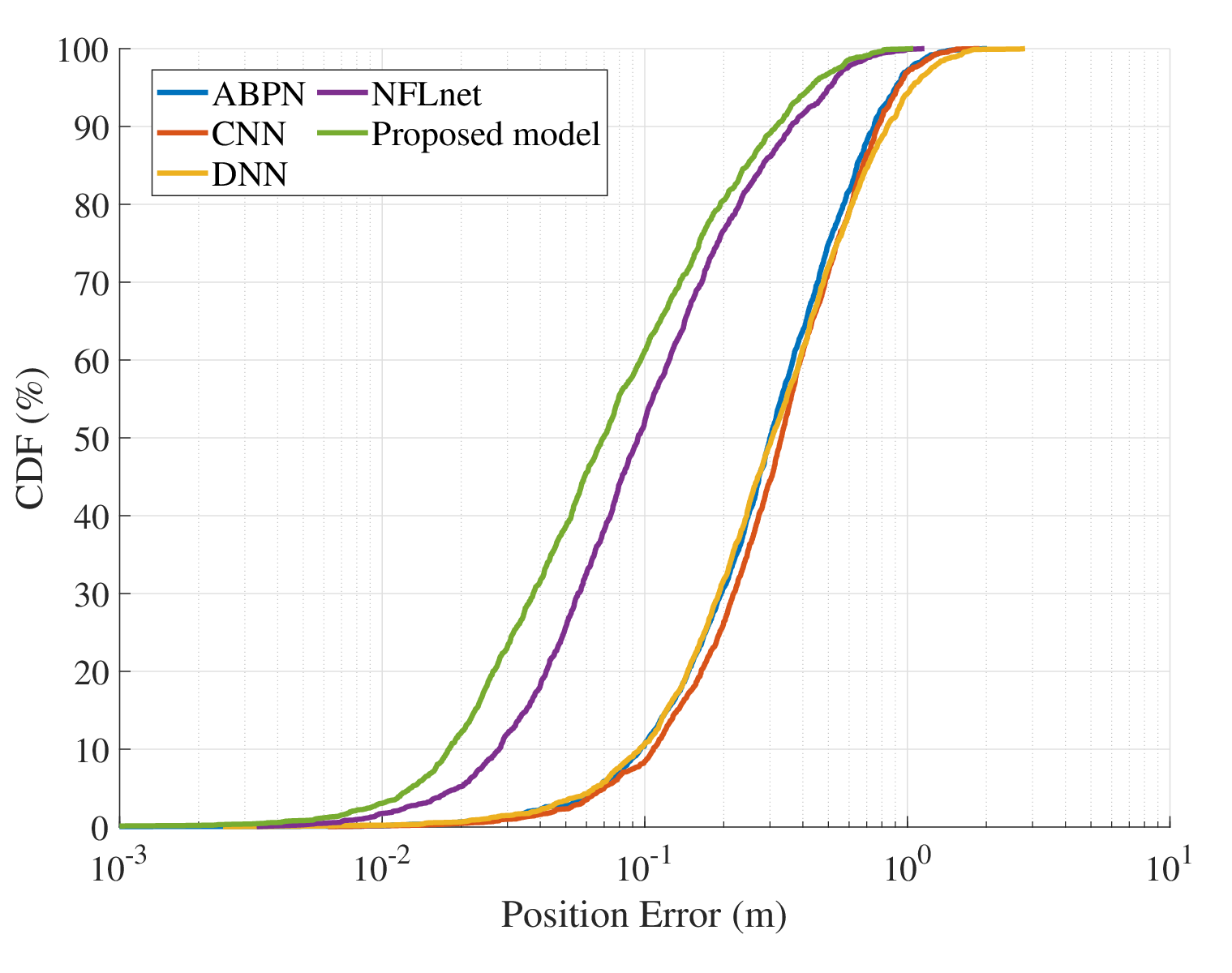}
        \label{CDF_50snap_0dB}
    }  
    \subfigure[Covariance matrix (50 snapshots) as input at 20 dB.]{
        \includegraphics[width=0.48\linewidth]{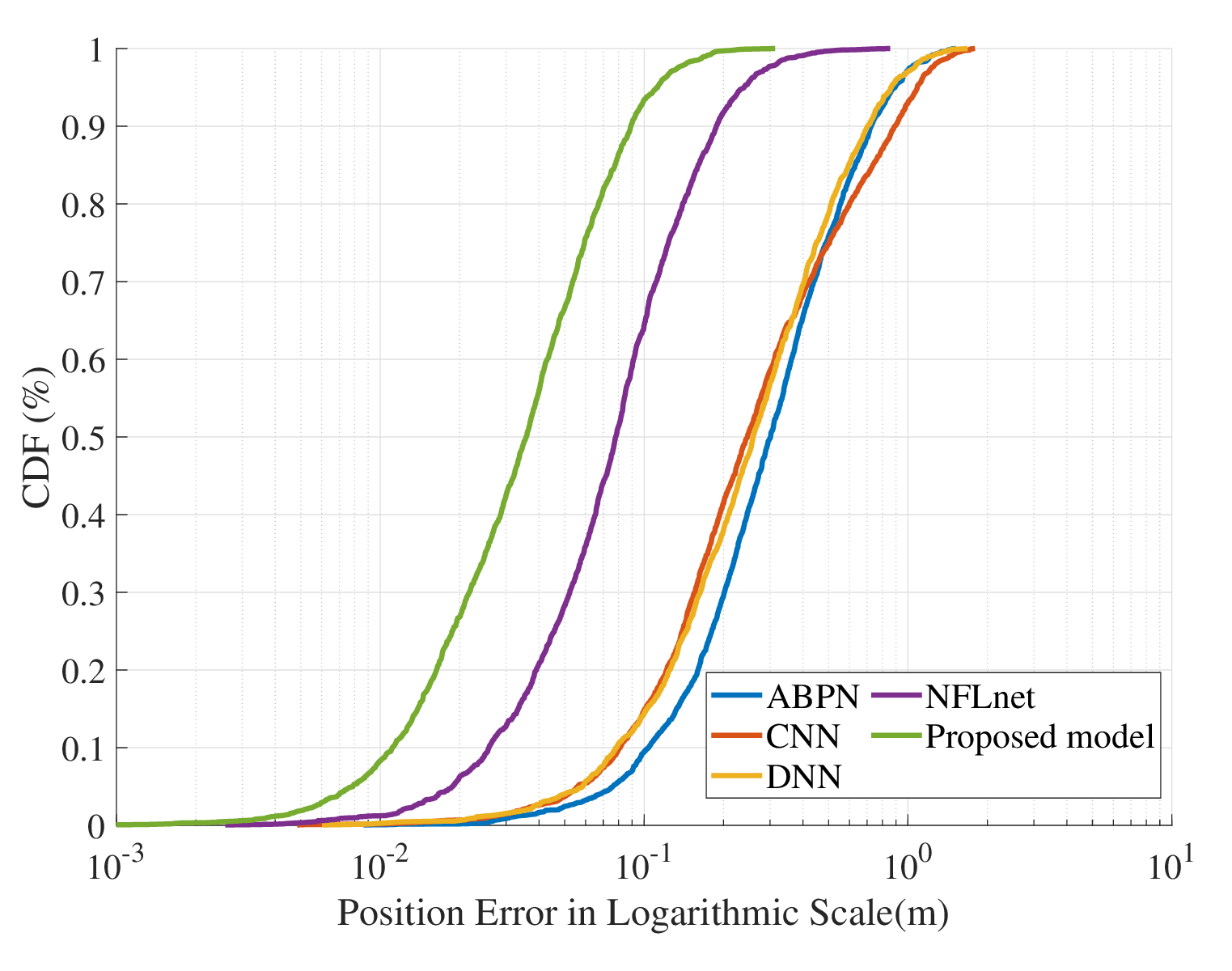}
        \label{CDF_50snap_20dB}
    }  
    \subfigure[Covariance matrix (100 snapshots) as input at 0 dB.]{
        \includegraphics[width=0.48\linewidth]{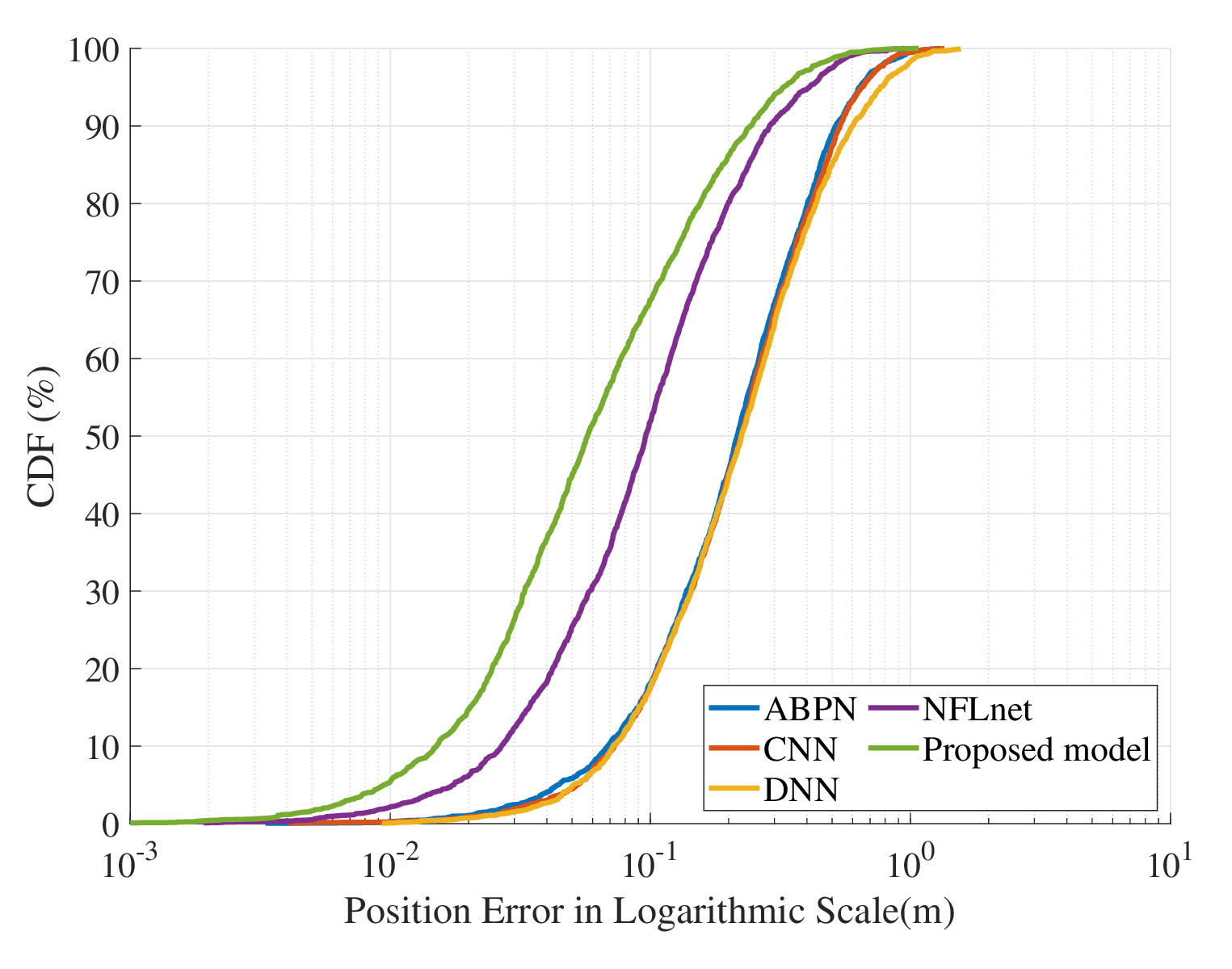}
        \label{CDF_100snap_0dB}
    }  
    \subfigure[Covariance matrix (100 snapshots) as input at 20 dB.]{
        \includegraphics[width=0.48\linewidth]{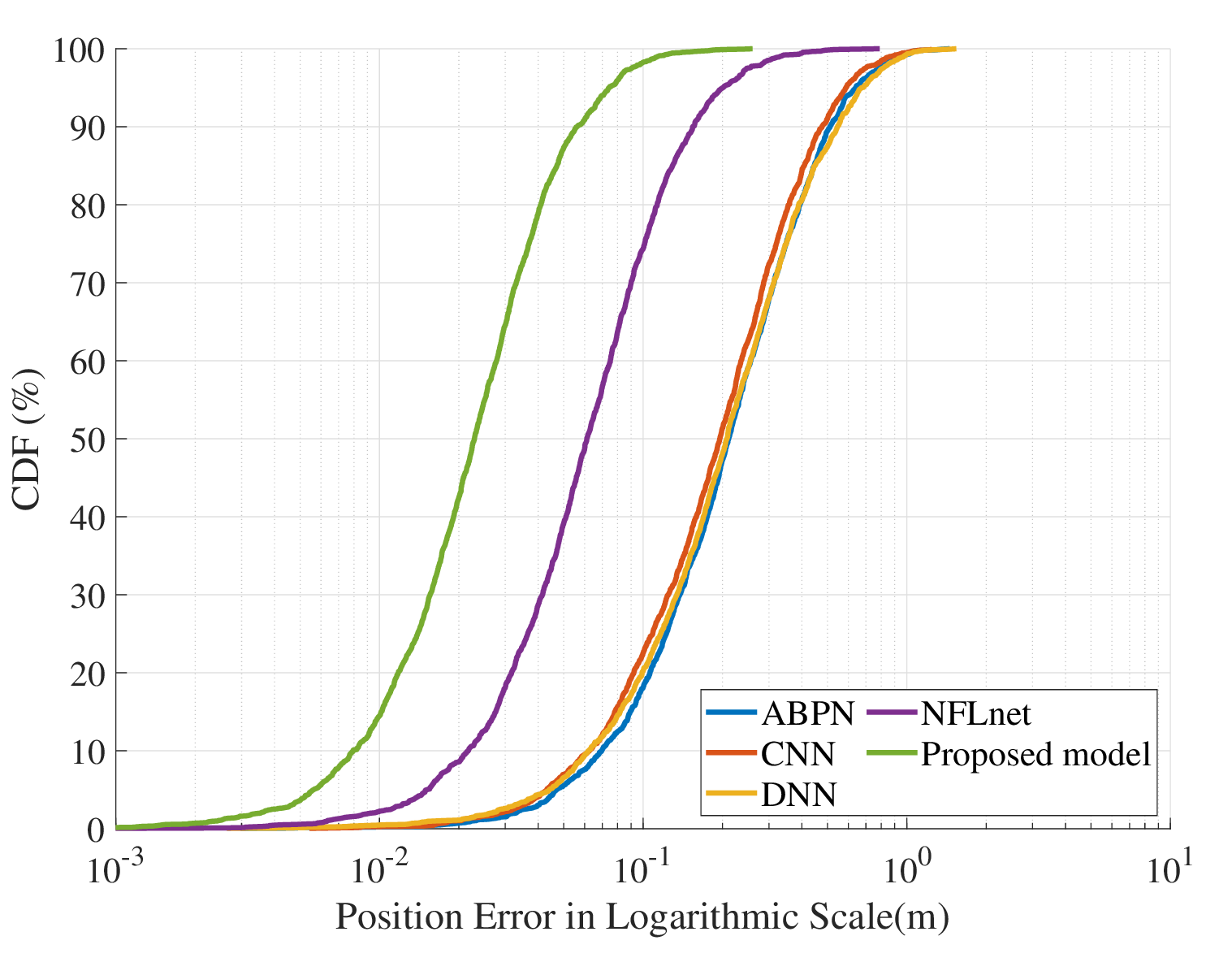}
        \label{CDF_100snap_20dB}
    }  
    \caption{The CDF of the positioning error of the proposed model, ABPN \cite{zhang2024deep}, CNN, MLP and NFLnet \cite{zhao2024nflnet} with various types of input and at different SNR levels.}
    \label{fig:CDF}
\end{figure*}

\begin{figure*}[t]
    \centering
    \subfigure[Mean positioning error at 0 dB.]{
        \includegraphics[width=0.48\linewidth]{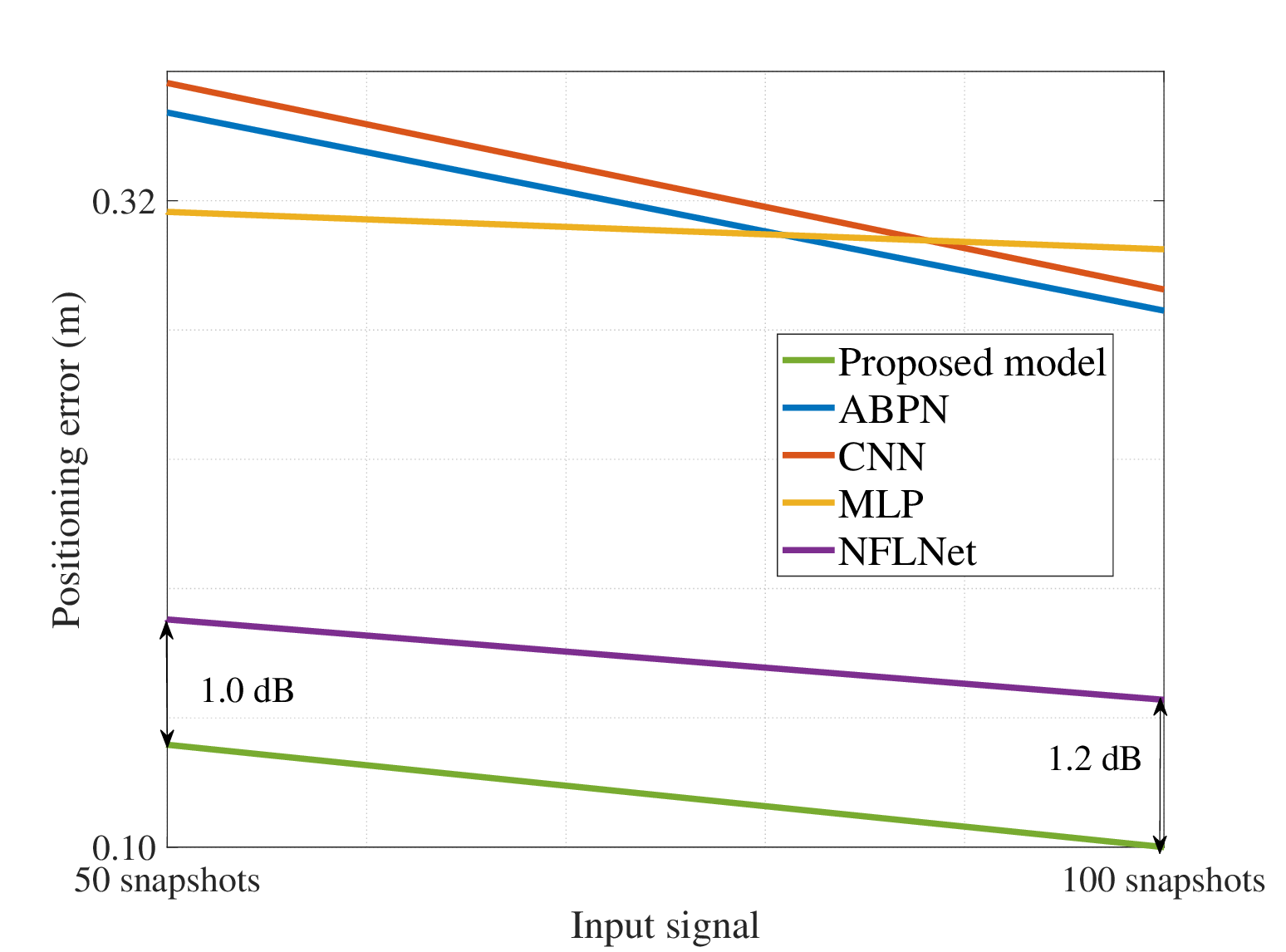}
        \label{mean_0dB}
    }
    \subfigure[Mean positioning error at 20 dB.]{
        \includegraphics[width=0.48\linewidth]{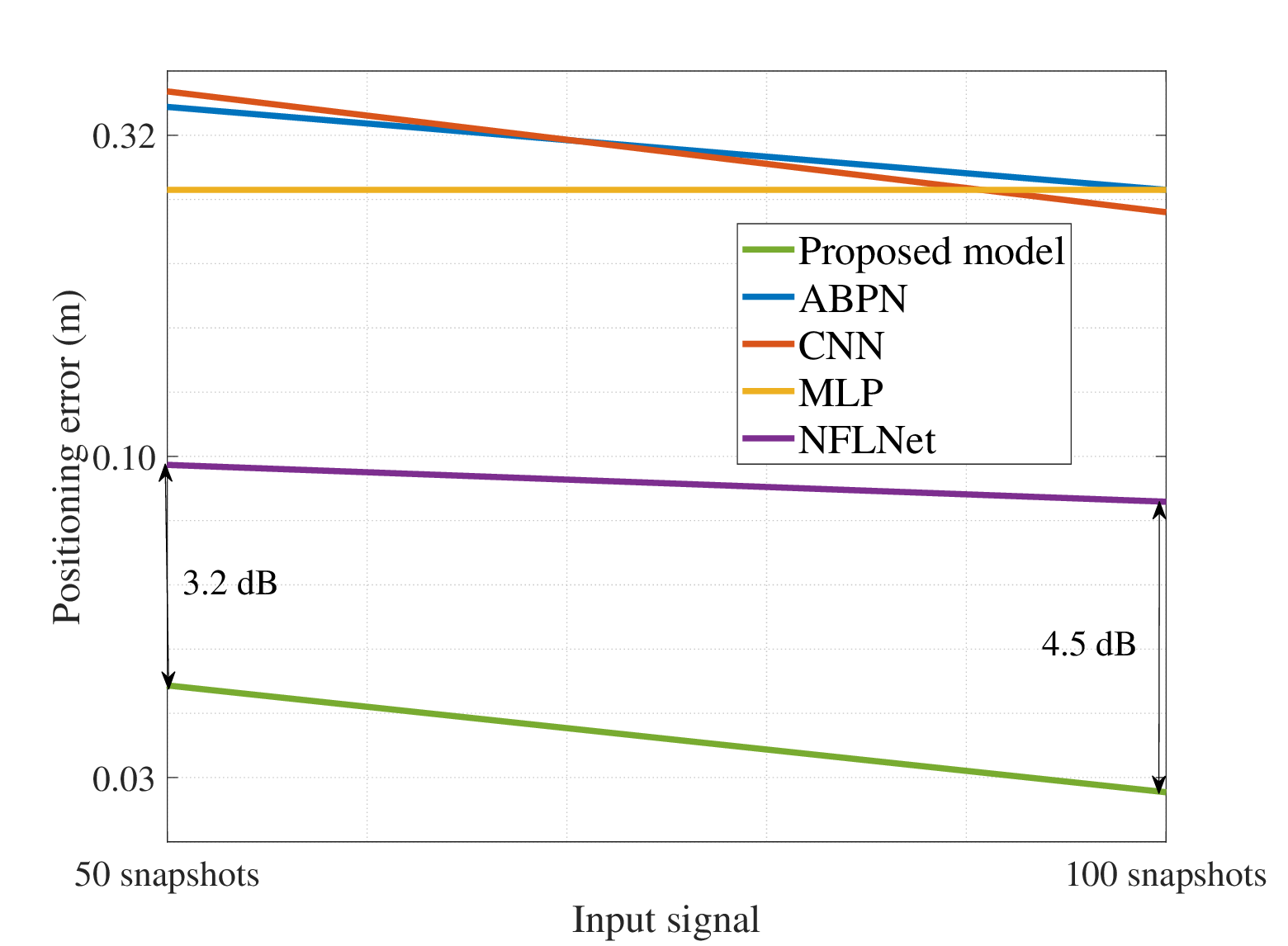}
        \label{mean_20dB}
    }
    \subfigure[Median positioning error at 0 dB.]{
        \includegraphics[width=0.48\linewidth]{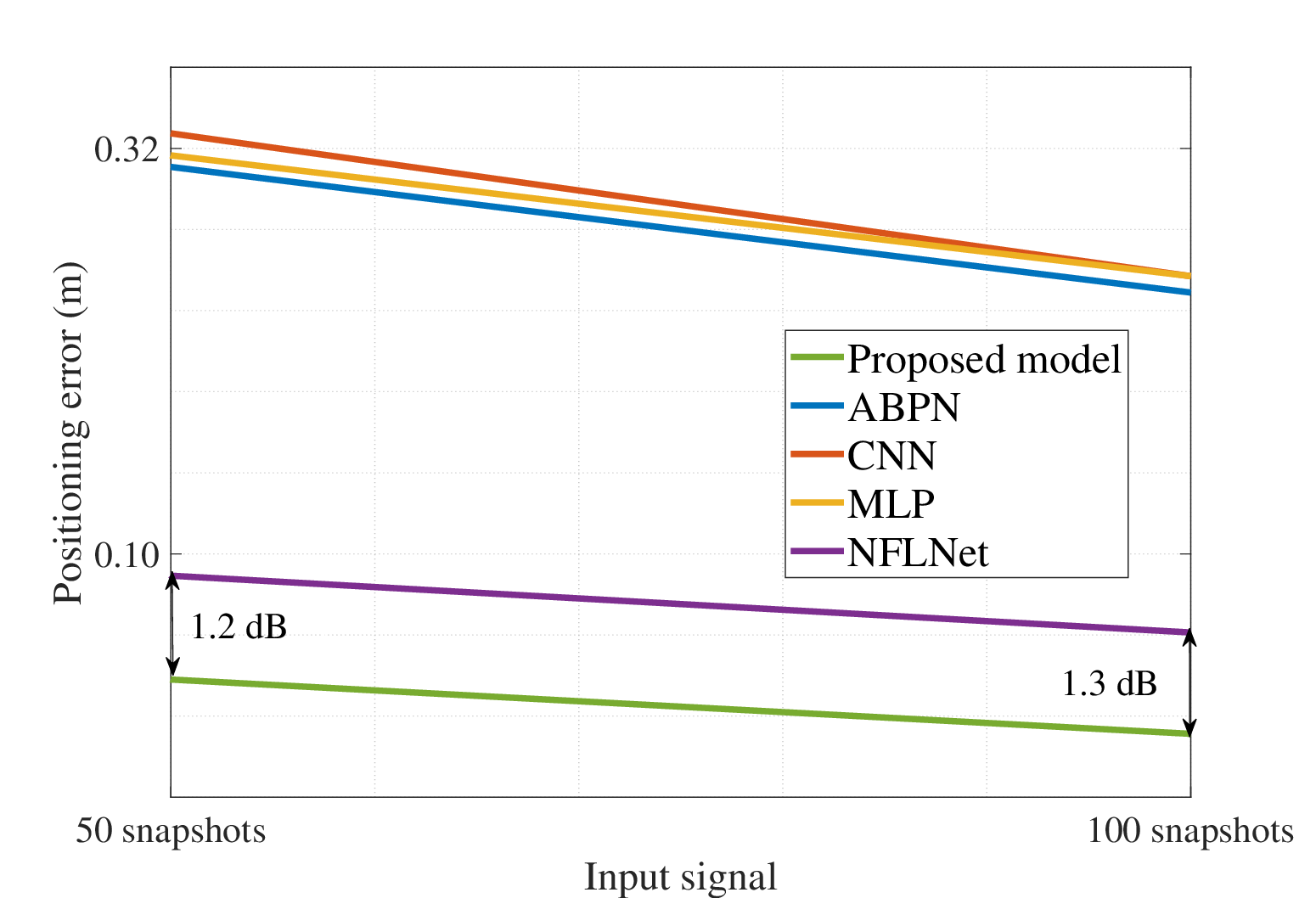}
        \label{median_0dB}
    }  
    \subfigure[Median positioning error at 20 dB.]{
        \includegraphics[width=0.48\linewidth]{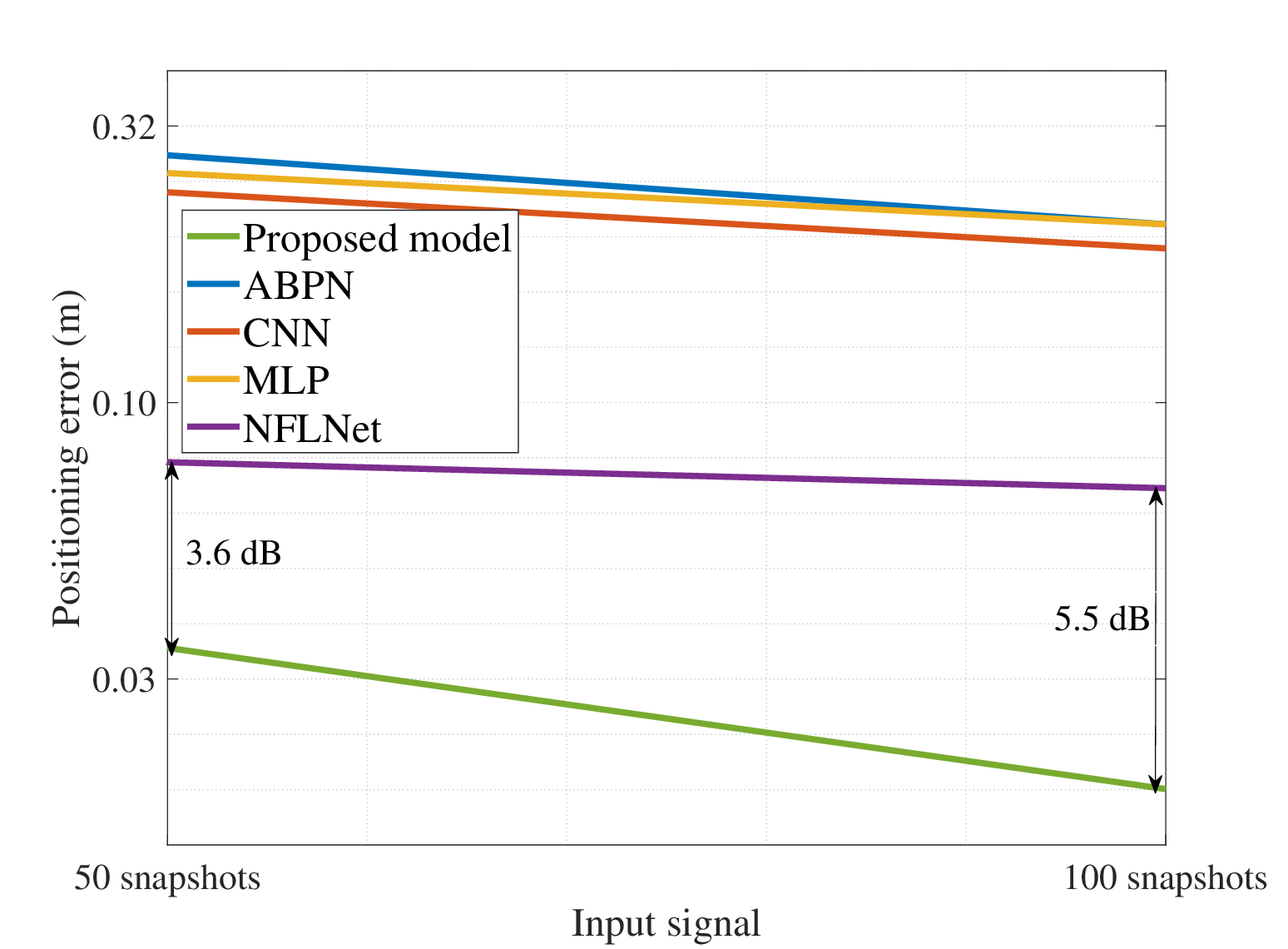}
        \label{median_20dB}
    }  
    \caption{Comparison between the proposed model, ABPN \cite{zhang2024deep}, CNN, MLP and NFLnet \cite{zhao2024nflnet} in terms of mean and median positioning error at 0 and 20 dB.}
    \label{fig:3x3_grid}
\end{figure*}

\begin{figure}[t]
    \centering
    \subfigure[Mean positioning error.]{
        \includegraphics[width=1\linewidth]{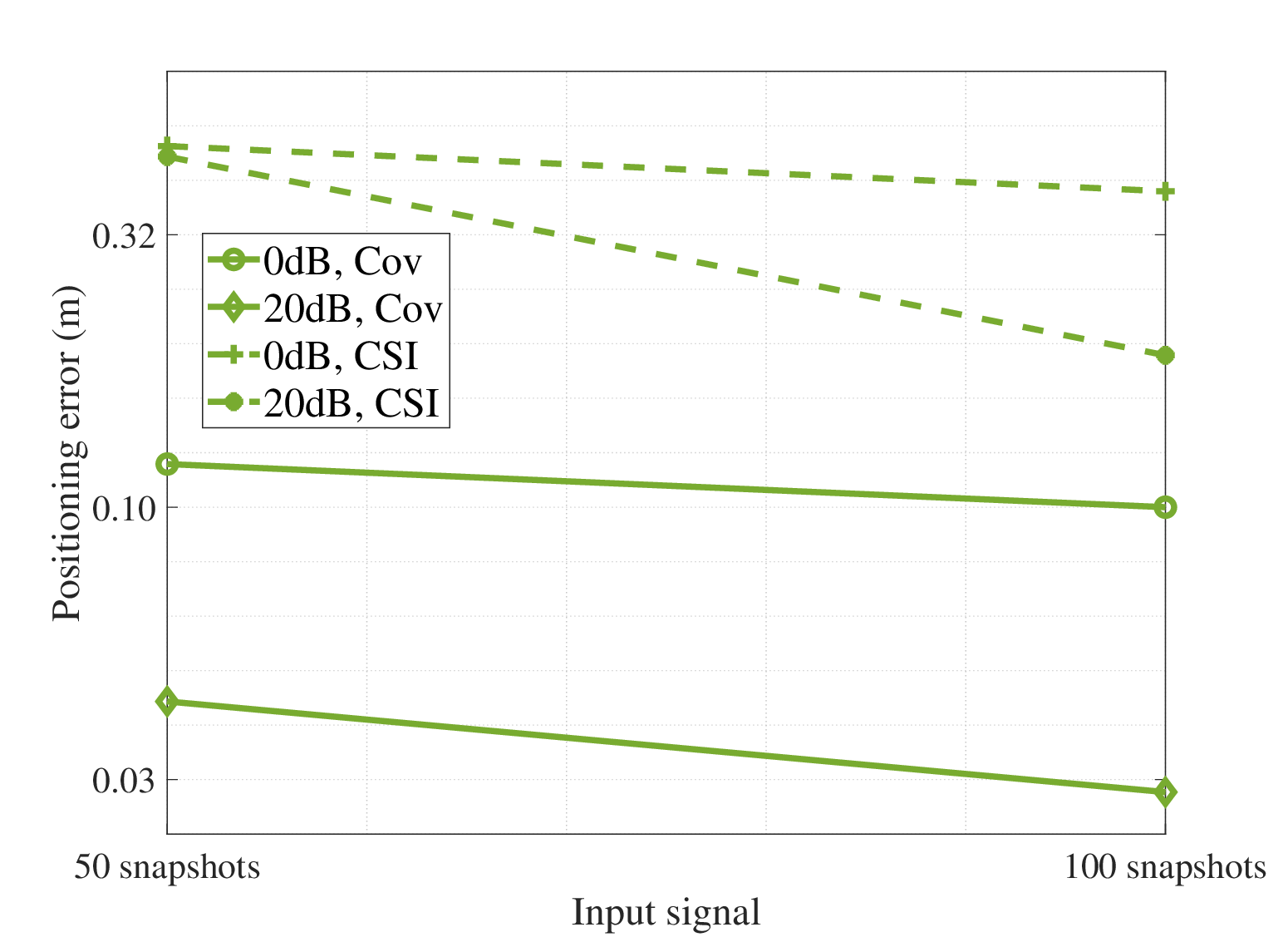}
        \label{fig:proposed_model_mean}
    }
    \subfigure[Median positioning error.]{
        \includegraphics[width=1\linewidth]{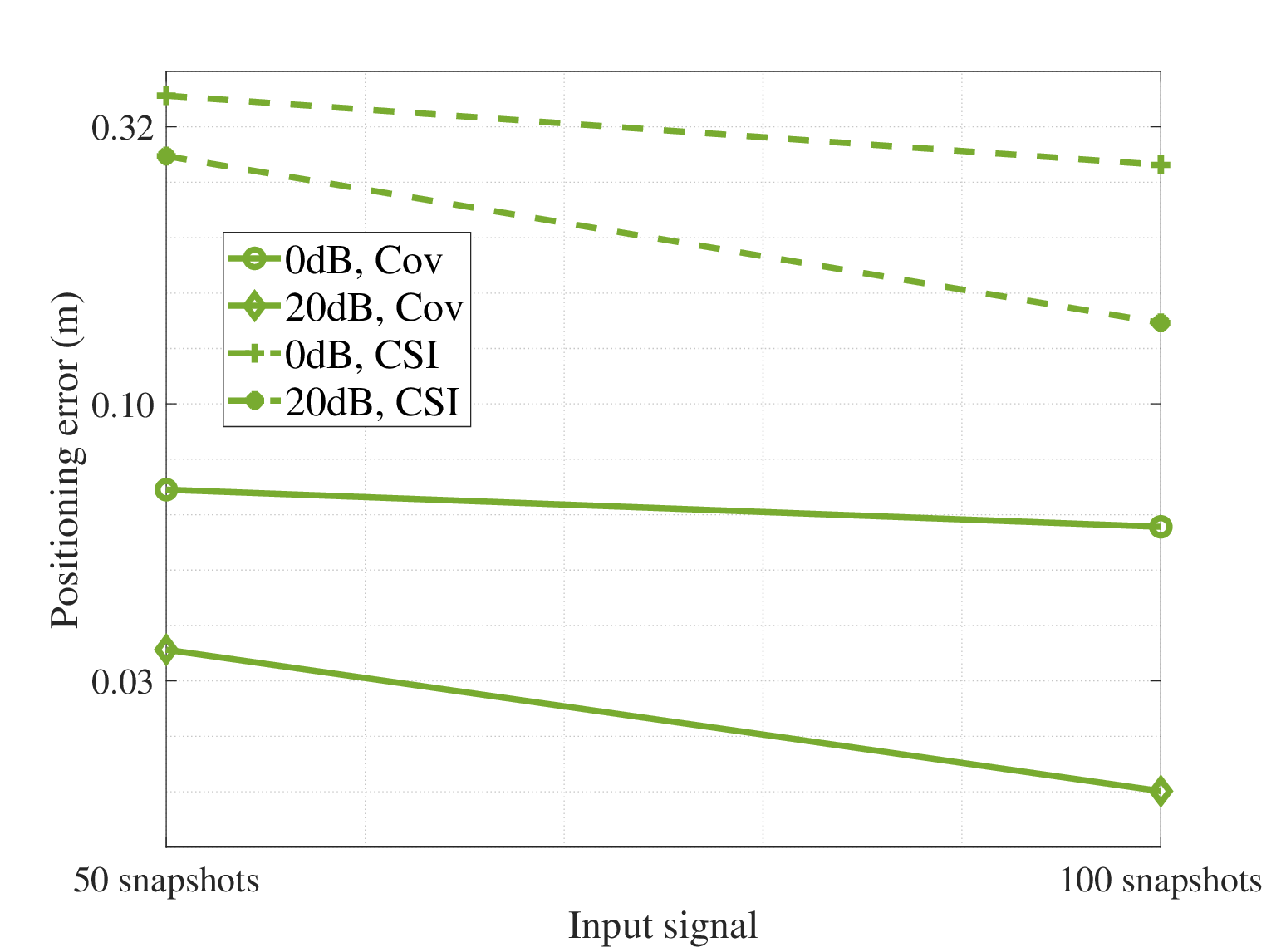}
        \label{fig:proposed_model_median}
    }
    \caption{Mean and median positioning error of the proposed model at 0 and 20 dB with the input signal as covariance metric and CSI.}
    \label{fig:proposed_model}
\end{figure}

The training and inference workflow of the proposed positioning model is described as follows. During the training phase, 8,000 training datasets each containing a $64\times 64$ covariance matrix $\mathbf{R}_{cov} \in \mathbb{R}^{64\times 64} $ paired with its corresponding distance-angle label $y= \left ( r,\eta  \right ) \in \mathbb{R} ^{2} $ are fed into the network after preprocessing. Through forward propagation, the CA and SA blocks collaboratively extract discriminative features. The positioning error is computed via a mean squared error (MSE) loss function \cite{wang2009mean,kim2021comparing}:
\begin{equation}
    \mathrm {MSE}= \frac{1}{N}\sum_{i= 1}^{N} \left [ \left ( \hat{r}^{\left ( i \right ) }- r^{\left ( i \right ) }_{true}\right)^{2}+ \left ( \hat{\eta }^{\left ( i\right) }- \eta ^{\left(i\right ) }_{true} \right )^{2}\right],
\end{equation}where $\hat{r}^{\left ( i \right ) }$ denotes the i-th predicted distance and $r^{\left ( i \right ) }_{true}$ denotes the i-th true distance, $\hat{\eta }^{\left ( i\right) }$ denotes the i-th predicted angle and $\eta ^{\left(i\right ) }_{true}  $ denotes the i-th true angle.

Network parameters are iteratively updated via backpropagation using the Adam optimizer \cite{bock2019proof,dereich2024convergence} with learning rate$= 3\times 10^{-4} $. During inference, 2,000 datasets are used for testing. $R_{cov} $ are directly processed by the trained model to generate sub-meter accuracy position estimates. The model achieves 22.87MB memory footprint with ultra-low 0.7ms inference latency per sample, meeting stringent real-time requirements for high-performance indoor positioning systems.

\section{Simulation results}

\subsection{Simulation settings}
To evaluate the effectiveness of the proposed model in near-field positioning, extensive simulations were conducted. The simulation environment emulates a near-field scenario by modeling the channel frequency response of a circular antenna array using MATLAB. CSI samples were generated under varying SNRs and numbers of snapshot. The covariance matrix, obtained through preprocessing of CSI data, serves as the model input and is paired with corresponding positional labels in a one-to-one manner. The circular array configuration comprises 64 antennas operating at a carrier frequency of 3.5 GHz, with elements distributed along a 1-meter-radius UCA. The angle $\eta $ and distance $r$ of the source were randomly sampled from the intervals $\left ( 30^{\circ},150^{\circ} \right ) $ and $\left ( 2m,10m \right ) $, respectively. Each simulation generated 10,000 samples. To ensure unbiased model evaluation, the dataset was randomly split into training and test sets (80\% for training and 20\% for testing). Table I shows the parameters of each layer of the proposed model structure. The proposed model was implemented in PyTorch and trained on NVIDIA GeForce RTX 4070 GPU with a batch size of 32 over 200 epochs. During the inference phase, the performance of the model was quantified using the root mean square error (RMSE) metric.
To comprehensively demonstrate the superiority of the proposed model in near-field positioning, we compare it with 4 representative AI-based models, namely ABPN \cite{zhang2024deep}, CNN, MLP and NFLnet \cite{zhao2024nflnet}. Specifically, ABPN utilizes both channel and spatial level attention mechanism to choose the most significant channel-level and pixel-level information of positioning. MLP comprises several fully-connected neural layers that perform non-linear mapping from covariance metric to position coordinates.

\subsection{Results analysis}
The positioning error of the proposed model and benchmarks are first compared in terms of cumulative distribution function (CDF) in Fig. \ref{fig:CDF}. Overall, the proposed model outperforms the benchmark models in all the scenarios of interest. By comparing the results between Fig. \ref{CDF_50snap_0dB} and Fig. \ref{CDF_50snap_20dB}, as well as between Fig. \ref{CDF_100snap_0dB} and Fig. \ref{CDF_100snap_20dB}, it can be observed that, at a fixed number of snapshots, the performance of the proposed model is significantly improved at higher SNR. In addition, the performance gap between the proposed model and the best benchmark model, i.e., NFLnet \cite{zhao2024nflnet}, at 20 dB is significantly larger than that at 0 dB. Apart from NFLnet, the remaining three benchmark models show similar positioning performance at 0 dB with 50 and 100 input snapshots. Additionally, comparing the results between Fig. \ref{CDF_50snap_0dB} and Fig. \ref{CDF_100snap_0dB}, or between \ref{CDF_50snap_20dB} and \ref{CDF_100snap_20dB}, at a given SNR, increasing the number of snapshots can effectively improve the positioning performance of all the models. 

We further compare the proposed model and benchmarks quantitatively in terms of mean positioning error and median positioning error in Fig. \ref{fig:3x3_grid}. The proposed model achieves the best performance, NFLnet ranks second, and the remaining three models exhibit similar performance. As illustrated in Fig. \ref{mean_0dB} and \ref{median_0dB}, at an SNR of 0 dB, the performance gap between the proposed model and NFLNet with 50 snapshots of covariance metric in terms of mean positioning error and median positioning error is 1.0 and 1.2 dB, respectively. By using 100 snapshots of the covariance metric, the performance gap increases slightly to 1.2 and 1.3 dB, respectively. As demonstrated in Fig. \ref{mean_20dB} and \ref{median_20dB}, the performance gap between the proposed model and NFLNet increases dramatically at SNR of 20 dB. This indicates that the proposed model can utilize the covariance metric of the input signal for positioning more effectively than the existing benchmark models. 

We further compare the performance of positioning using covariance metric and CSI in Fig. \ref{fig:proposed_model}. A dramatic performance gap between the input of covariance metric and CSI is observed for both the mean positioning error and median positioning error in Fig. \ref{fig:proposed_model_mean} and \ref{fig:proposed_model_median}, respectively. The performance gap is over 7 dB, which indicates that a much higher positioning accuracy can be achieved by the proposed model using the covariance metric as the input, which only requires a straightforward processing given in Eq. \ref{eq:cov}. In addition, a dedicated model must be trained for CSI inputs with different numbers of snapshots, indicating that CSI-based input is inefficient in terms of model training and adaptation.
\section{Conclusion}
In 6G mobile communication systems, XL-MIMO is progressively replacing MIMO systems as a new research focus, showing significant potential in positioning. However, the near-field effect causes challenges in the design of the positioning algorithm. In this paper, we propose an attention-enhanced deep learning model for positioning. Specifically, we develop a dual-path channel attention mechanism alongside a spatial attention mechanism. These mechanisms effectively integrate channel-level and spatial-level features by leveraging parallel average pooling and max pooling techniques, facilitating comprehensive feature extraction. The simulation results illustrate that the proposed attention-enhanced deep learning method significantly exceeds existing benchmarks. The proposed model achieves superior positioning accuracy by utilizing covariance metrics of the input signal. Moreover, simulation results reveal that covariance metric is advantageous for positioning over CSI in terms of positioning accuracy and model efficiency.

\ifCLASSOPTIONcaptionsoff
  \newpage
\fi

\bibliography{reference.bib}
\end{document}